\documentstyle[twoside,psfig]{article}

\catcode`\@=11
\long\def\@makefntext#1{
\protect\noindent \hbox to 3.2pt {\hskip-.9pt  
$^{{\eightrm\@thefnmark}}$\hfil}#1\hfill}               

\def\thefootnote{\fnsymbol{footnote}}
\def\@makefnmark{\hbox to 0pt{$^{\@thefnmark}$\hss}}    
        
\def\ps@myheadings{\let\@mkboth\@gobbletwo
\def\@oddhead{\hbox{}
\rightmark\hfil\eightrm\thepage}   
\def\@oddfoot{}\def\@evenhead{\eightrm\thepage\hfil
\leftmark\hbox{}}\def\@evenfoot{}
\def\sectionmark##1{}\def\subsectionmark##1{}}

\oddsidemargin=\evensidemargin
\renewcommand{\thefootnote}{\fnsymbol{footnote}}

\newcounter{sectionc}\newcounter{subsectionc}\newcounter{subsubsectionc}
\renewcommand{\section}[1] {\vspace{12pt}\addtocounter{sectionc}{1} 
\setcounter{subsectionc}{0}\setcounter{subsubsectionc}{0}\noindent 
        {\tenbf\thesectionc. #1}\par\vspace{5pt}}
\renewcommand{\subsection}[1] {\vspace{12pt}\addtocounter{subsectionc}{1} 
        \setcounter{subsubsectionc}{0}\noindent 
        {\bf\thesectionc.\thesubsectionc. {\kern1pt \bfit #1}}\par\vspace{5pt}}
\renewcommand{\subsubsection}[1] {\vspace{12pt}\addtocounter{subsubsectionc}{1}
        \noindent{\tenrm\thesectionc.\thesubsectionc.\thesubsubsectionc.
        {\kern1pt \tenit #1}}\par\vspace{5pt}}
\newcommand{\nonumsection}[1] {\vspace{12pt}\noindent{\tenbf #1}
        \par\vspace{5pt}}

\newcounter{appendixc}
\newcounter{subappendixc}[appendixc]
\newcounter{subsubappendixc}[subappendixc]
\renewcommand{\thesubappendixc}{\Alph{appendixc}.\arabic{subappendixc}}
\renewcommand{\thesubsubappendixc}
        {\Alph{appendixc}.\arabic{subappendixc}.\arabic{subsubappendixc}}

\renewcommand{\appendix}[1] {\vspace{12pt}
        \refstepcounter{appendixc}
        \setcounter{figure}{0}
        \setcounter{table}{0}
        \setcounter{lemma}{0}
        \setcounter{theorem}{0}
        \setcounter{corollary}{0}
        \setcounter{definition}{0}
        \setcounter{equation}{0}
        \renewcommand{\thefigure}{\Alph{appendixc}.\arabic{figure}}
        \renewcommand{\thetable}{\Alph{appendixc}.\arabic{table}}
        \renewcommand{\theappendixc}{\Alph{appendixc}}
        \renewcommand{\thelemma}{\Alph{appendixc}.\arabic{lemma}}
        \renewcommand{\thetheorem}{\Alph{appendixc}.\arabic{theorem}}
        \renewcommand{\thedefinition}{\Alph{appendixc}.\arabic{definition}}
        \renewcommand{\thecorollary}{\Alph{appendixc}.\arabic{corollary}}
        \renewcommand{\theequation}{\Alph{appendixc}.\arabic{equation}}
        \noindent{\tenbf Appendix \theappendixc #1}\par\vspace{5pt}}
\newcommand{\subappendix}[1] {\vspace{12pt}
        \refstepcounter{subappendixc}
        \noindent{\bf Appendix \thesubappendixc. {\kern1pt \bfit #1}}
        \par\vspace{5pt}}
\newcommand{\subsubappendix}[1] {\vspace{12pt}
        \refstepcounter{subsubappendixc}
        \noindent{\rm Appendix \thesubsubappendixc. {\kern1pt \tenit #1}}
        \par\vspace{5pt}}

\newcommand{\textlineskip}{\baselineskip=13pt}
\newcommand{\smalllineskip}{\baselineskip=10pt}

\def\eightcirc{
\begin{picture}(0,0)
\put(4.4,1.8){\circle{6.5}}
\end{picture}}
\def\eightcopyright{\eightcirc\kern2.7pt\hbox{\eightrm c}}

\def\abstracts#1#2#3{{
        \centering{\begin{minipage}{4.5in}\footnotesize\baselineskip=10pt
        \parindent=0pt #1\par 
        \parindent=15pt #2\par
        \parindent=15pt #3
        \end{minipage}}\par}} 

\newcommand{\bibit}{\nineit}
\newcommand{\bibbf}{\ninebf}
\renewenvironment{thebibliography}[1]
        {\frenchspacing
         \ninerm\baselineskip=11pt
         \begin{list}{\arabic{enumi}.}
        {\usecounter{enumi}\setlength{\parsep}{0pt}     
         \setlength{\leftmargin 12.7pt}{\rightmargin 0pt} 
         \setlength{\itemsep}{0pt} \settowidth
        {\labelwidth}{#1.}\sloppy}}{\end{list}}

\newcounter{itemlistc}
\newcounter{romanlistc}
\newcounter{alphlistc}
\newcounter{arabiclistc}

\newcommand{\fcaption}[1]{
        \refstepcounter{figure}
        \setbox\@tempboxa = \hbox{\footnotesize Fig.~\thefigure. #1}
        \ifdim \wd\@tempboxa > 5in
           {\begin{center}
        \parbox{5in}{\footnotesize\smalllineskip Fig.~\thefigure. #1}
            \end{center}}
        \else
             {\begin{center}
             {\footnotesize Fig.~\thefigure. #1}
              \end{center}}
        \fi}

\newcommand{\tcaption}[1]{
        \refstepcounter{table}
        \setbox\@tempboxa = \hbox{\footnotesize Table~\thetable. #1}
        \ifdim \wd\@tempboxa > 5in
           {\begin{center}
        \parbox{5in}{\footnotesize\smalllineskip Table~\thetable. #1}
            \end{center}}
        \else
             {\begin{center}
             {\footnotesize Table~\thetable. #1}
              \end{center}}
        \fi}

\def\@citex[#1]#2{\if@filesw\immediate\write\@auxout
        {\string\citation{#2}}\fi
\def\@citea{}\@cite{\@for\@citeb:=#2\do
        {\@citea\def\@citea{,}\@ifundefined
        {b@\@citeb}{{\bf ?}\@warning
        {Citation `\@citeb' on page \thepage \space undefined}}
        {\csname b@\@citeb\endcsname}}}{#1}}

\newif\if@cghi
\def\cite{\@cghitrue\@ifnextchar [{\@tempswatrue
        \@citex}{\@tempswafalse\@citex[]}}
\def\citelow{\@cghifalse\@ifnextchar [{\@tempswatrue
        \@citex}{\@tempswafalse\@citex[]}}
\def\@cite#1#2{{$\null^{#1}$\if@tempswa\typeout
        {IJCGA warning: optional citation argument 
        ignored: `#2'} \fi}}

\def\pmb#1{\setbox0=\hbox{#1}
        \kern-.025em\copy0\kern-\wd0
        \kern.05em\copy0\kern-\wd0
        \kern-.025em\raise.0433em\box0}

\def\fnt#1#2{\footnotetext{\kern-.3em
        {$^{\mbox{\scriptsize #1}}$}{#2}}}

\def\fpage#1{\begingroup
\voffset=.3in
\thispagestyle{empty}\begin{table}[b]\centerline{\footnotesize #1}
        \end{table}\endgroup}

\def\runninghead#1#2{\pagestyle{myheadings}
\markboth{{\protect\footnotesize\it{\quad #1}}\hfill}
{\hfill{\protect\footnotesize\it{#2\quad}}}}
\font\tenrm=cmr10
\font\tenit=cmti10 
\font\tenbf=cmbx10
\font\bfit=cmbxti10 at 10pt
\font\ninerm=cmr9
\font\nineit=cmti9
\font\ninebf=cmbx9
\font\eightrm=cmr8

\def\qed{\hbox{${\vcenter{\vbox{                        
   \hrule height 0.4pt\hbox{\vrule width 0.4pt height 6pt
   \kern5pt\vrule width 0.4pt}\hrule height 0.4pt}}}$}}

\renewcommand{\thefootnote}{\fnsymbol{footnote}}        

\begin{document}
\setlength{\textheight}{7.7truein}  

\runninghead{D. Ebert, R. N. Faustov and V. O. Galkin}{Decay Constants
  of Heavy-Light Mesons}

\normalsize\textlineskip
\thispagestyle{empty}
\setcounter{page}{1}
\begin{flushright}\normalsize HU-EP-02/15\\
\vskip 1cm
\end{flushright}

\vspace*{0.88truein}

\fpage{1}
\centerline{\bf DECAY CONSTANTS OF HEAVY-LIGHT MESONS  }
\baselineskip=13pt
\centerline{\bf IN THE RELATIVISTIC QUARK MODEL}
\vspace*{0.37truein}
\centerline{\footnotesize D. EBERT}
\baselineskip=12pt
\centerline{\footnotesize\it Institut f\"ur Physik,
  Humboldt--Universit\"at zu Berlin, Invalidenstr.110, D-10115 Berlin, Germany}
\vspace*{10pt}

\centerline{\footnotesize R. N. FAUSTOV and V. O. GALKIN}
\baselineskip=12pt
\centerline{\footnotesize\it Institut f\"ur Physik,
  Humboldt--Universit\"at zu Berlin, Invalidenstr.110, D-10115 Berlin,
  Germany} 
\baselineskip=10pt
\centerline{\footnotesize\it and}
\baselineskip=10pt
\centerline{\footnotesize\it Russian Academy of Sciences, Scientific Council for
Cybernetics, Vavilov Str. 40,}
\baselineskip=10pt
\centerline{\footnotesize\it  Moscow 117333, Russia}


\vspace*{0.21truein}
\abstracts{The decay constants of pseudoscalar and vector heavy-light
  mesons are calculated in the framework of the relativistic quark
  model with the completely relativistic treatment of the light quark. It
  is argued that relativistic effects play a significant
  role. Good agreement of the model predictions with recent lattice
  and QCD sum rule calculations as well as available experimental data
  is found.}{}{}

\vspace*{10pt}

\textlineskip                  
\vspace*{12pt}                 

\setcounter{footnote}{0}
\renewcommand{\thefootnote}{\alph{footnote}}

\noindent
The pseudoscalar and vector decay constants of the heavy-light mesons
are important hadronic characteristics of these
mesons. They determine such quantities as the weak
decay rates of mesons to lepton pairs and the
magnitude of $B_d-\bar B_d$ and $B_s-\bar B_s$ mixings. These constants
also enter the nonleptonic decay rates, considered in the
factorization approximation. In this brief note we consider them in the
framework of the relativistic quark model.

The decay constants $f_P$ and $f_V$ of the pseudoscalar ($P$) and
vector ($V$) mesons parameterize the matrix elements of the weak
current between the corresponding meson and the vacuum. In the case of
a meson, composed from a heavy antiquark $\bar Q$ and a light quark
$q$, they are defined by 
\begin{eqnarray}
  \label{eq:dc}
  \left<0|\bar Q \gamma^\mu\gamma_5 q|P({\bf K})\right>&=& i f_P
  K^\mu,\\ \cr
\left<0|\bar Q \gamma^\mu q|V({\bf K},\varepsilon)\right>&=& f_V
  M_V \varepsilon^\mu,
\end{eqnarray}
where ${\bf K}$ is the meson momentum, $\varepsilon$ and $M_V$ are
the polarisation vector and mass of the vector meson.

In the relativistic quark model the decay constants can be expressed
through the meson wave function $\Phi_{P,V}(p)$ in the momentum space
and are given by\cite{gfm,g}
\begin{eqnarray}
  \label{eq:fpv}
  f_{P,V}&=&\sqrt{\frac{12}{M_{P,V}}}\int \frac{d^3
  p}{(2\pi)^3}\left(\frac{\epsilon_q(p)+m_q}{2\epsilon_q(p)}\right)^{1/2}
  \left(\frac{\epsilon_Q(p)+m_Q}{2\epsilon_Q(p)}\right)^{1/2}
  \nonumber\\ \cr
&&\times \left\{ 1
  +\lambda_{P,V}\,\frac{{\bf p}^2}{[\epsilon_q(p)+m_q][\epsilon_Q(p)+m_Q]}\right\}
  \Phi_{P,V}(p),
\end{eqnarray}
with $\lambda_P=-1$ and $\lambda_V=1/3$.
In the nonrelativistic limit $p^2/m^2\to 0$ these expressions for
decay constants give the well-known formula
\begin{equation}
\label{eq:fnr}
f_P^{\rm NR}=f_V^{\rm NR}=
\sqrt{\frac{12}{M_{P,V}}}\left|\Psi_{P,V}(0)\right|,
\end{equation}
where $\Psi_{P,V}(0)$ is the meson wave function at the origin
$r=0$. In the case of  the heavy-light $B$ and $D$
mesons, considered in the infinitely heavy quark mass
limit $m_Q\to \infty$, the meson wave function is determined by the
light quark only. Therefore in this limit, as it follows from 
Eq.~(\ref{eq:fpv}),  the decay constants scale inversely
proportional to the square root of the meson mass:
$f_P\sqrt{M_P}={\rm const}$. This leads to the relation $f_B^{\rm
  NR}/f_D^{\rm NR}=\sqrt{M_D/M_B}$  between $B$ and $D$
decay constants. However, the relativistic effects
break down this scale behaviour. Expanding expression
(\ref{eq:fpv}) in $1/m_Q$, we get
\begin{eqnarray}
  \label{eq:rf}
  \frac{f_B}{f_D}&=&\sqrt{\frac{M_D}{M_B}}\Biggl\{1+\frac12\left(\frac1{m_c}
  -\frac1{m_b}\right)\frac{\displaystyle \int \frac{d^3
  p}{(2\pi)^3}
  \frac{{\bf p}^2}{\sqrt{2\epsilon_q(p)[\epsilon_q(p)+m_q]}}
  \Phi_P(p) }{\displaystyle \int \frac{d^3
  p}{(2\pi)^3}\left(\frac{\epsilon_q(p)+m_q}{2\epsilon_q(p)}\right)^{1/2}
  \Phi_P(p)}\cr \cr\cr
&&\left.
+O\left(\frac1{m_Q^2}\right)\right\}
\approx \sqrt{\frac{M_D}{M_B}}\left\{1
+0.20 +O\left(\frac1{m_Q^2}\right)\right\}.
\end{eqnarray}
Thus we see that the first order corrections in $1/m_Q$ shift the ratio
of the pseudoscalar decay constants of $D$ and $B$ mesons by  about 
$20\%$. As a result, relativistic values of $f_B$ and  $f_D$ are 
closer to each other than nonrelativistic ones.  As it will be seen
later, this fact is important for bringing quark model predictions in
agreement with recent lattice and QCD sum rule calculations.  

\begin{table}[htbp]
\tcaption{Pseudoscalar and vector decay constants of mesons with open
  flavour (in MeV).}
\label{tab:1}
\begin{tabular}{ccccccccccc}\\
\hline
Meson& $D$ &$D^*$ &$D_s$&$D_s^*$ &$B$ &$B^*$&$B_s$&$B_s^*$
&$B_c$ &$B_c^*$\\
\hline
$f^{\rm NR}$&332 &332& 384& 384 & 213& 213& 278& 278& 562& 562\\ 
$f_{P,V}$& 243 &315& 266&  335& 178 & 195& 196& 214& 433& 503\\ 
\hline
\end{tabular}
\end{table}

The calculated values of the pseudoscalar
and vector decay constants of heavy-light mesons in our model are
displayed in Table~\ref{tab:1}. In the first row we show nonrelativistic
predictions and in the second row relativistic results
(\ref{eq:fpv}). For these calculations we used the wave functions
of the heavy-light mesons obtained in the relativistic quark
model based on the quasipotential approach in quantum field
theory in calculating their mass spectra\cite{egf}. It is
important to stress that the light quark was treated completely
relativistically, without employing an unjustified expansion in inverse
powers of the light quark mass. Only the heavy quark expansion was
used in order to simplify calculations. The relativistic treatment of
the light quark has a significant impact on the wave functions and, as
a result, on the decay constants of the heavy-light mesons. It 
increases the ratio $f_B/f_D$ from 0.64 in the nonrelativistic limit
to  0.74 and decreases the ratios $f_{B_s}/f_B$ from 1.3 to 1.1
and $f_{D_s}/f_D$ from 1.2 to 1.1, respectively.     

\begin{table}[htbp]
\tcaption{Pseudoscalar decay constants of the heavy-light mesons (in
  MeV). Comparison of our results with averaged lattice data, QCD sum
  rule predictions and averaged experimental data.}
\label{tab:2}
\begin{tabular}{ccccccc}\\
\hline
Ref.& $f_B$& $f_{B_s}$&$f_{B_s}/f_B$& $f_D$& $f_{D_s}$&$f_{D_s}/f_D$\\ 
\hline
This work& 178(15) & 196(20) & 1.11& 243(25) & 266(25)& 1.10 \\ 
{\small\bf Lattice}\cite{r}& & & & & & \\
Quenched&173(23)& 200(20)& 1.15(3)& 203(14)& 230(14)& 1.12(2)\\
Unquenched& 198(30)& 230(30)&1.16(5)&226(15)&250(30)& 1.12(4)\\
{\small\bf QCD SR}& & & & & & \\
Ref.\cite{n}& 203(23) & 236(30)&1.16(4)&203(23)&235(24)&1.15(4)\\
Ref.\cite{ps}& 206(20)& & &195(20)& & \\
Ref.\cite{jl}& 210(19) &244(21)&1.16 & & & \\
$\!${\small\bf Experiment}& & & & & & \\
PDG\cite{pdg}& & & &&$\!\!\!$ 
280(19)(28)(34)$\!\!\!\!$& \\
Ref.\cite{sr}&  & && &$\!\!$ 264(15)(33)(5)$\!\!\!$& \\
\hline\\
\end{tabular}
\end{table}

In Table~\ref{tab:2} we compare our predictions\footnote{We roughly
  estimate the error in our calculations (mainly coming from the wave
  functions) to be about $10\%$.}  \ for pseudoscalar decay
constants of the heavy-light mesons as well as their ratios with other
theoretical results and
available experimental data. The pseudoscalar decay constants of
$B$ and $D$ mesons were studied rather extensively in different
theoretical approaches. In this Table we present the averaged lattice
data obtained both in quenched and unquenched approximations from the
recent review\cite{r}. We also list the most recent QCD sum rule
predictions\cite{n,ps,jl}. Reliable experimental data at present
exist only for $f_{D_s}$, which was measured by several experimental
collaborations (ALEPH, DELPHI, L3, OPAL, Beatrice, CLEO, E653, WA75,
BES) both in the $D_s\to\mu\nu_\mu$ and the $D_s\to\tau\nu_\tau$ decay
channels. Unfortunately, experimental errors are still large. In
Table~\ref{tab:2} we present two experimental averages: one from
PDG\cite{pdg} and the other from the recent experimental
review\cite{sr}. There exists also the first experimental value for
$f_{D^+}=300^{+180+80}_{-150-40}$~MeV, but based on only one
$D^+\to\mu^+\nu_\mu$ event\cite{bai} and thus with very large
errors. We see from Table~\ref{tab:2} that there is good (within error
bars) agreement
between all presented theoretical predictions as well as with
available experimental data. However, the QCD sum rules seem tend to
prefer for the $f_B/f_D$ ratio the value close to 1, while somewhat smaller
values are favoured by the lattice (0.85) and by our  model (0.74).         

We can also compare our prediction for the vector decay constant
$f_{B^*}$ with the recent lattice calculation in quenched
approximation\cite{bwd}: 
\[f_{B^*}=177\pm6\pm 17\ {\rm MeV\ \ \ and\ \ \ \ }
f_{B^*}/f_B=1.01\pm0.01{}^{+0.04}_{-0.01}.
\]  
The ratios of vector and pseudoscalar decay constants $f_V/f_P$ are
also predicted by heavy quark effective theory\cite{neub1} (HQET)
\[
f_{B^*}/f_B\cong 1.07\pm 0.03, \quad  f_{D^*}/f_D\cong 1.31\pm 0.08.
\]
For these ratios in our model we have
\[
f_{B^*}/f_B\cong 1.09, \quad  f_{D^*}/f_D\cong 1.30
\]
in good agreement with both lattice and HQET results.

In summary, we calculated the decay constants of the pseudoscalar and
vector heavy-light mesons in the framework of the relativistic quark
model paying special attention to the complete relativistic treatment
of the light quark. It was  found that the relativistic effects
significantly reduce the values of the decay constants and
violate the scaling behaviour with masses of the decay
constants which emerge in the nonrelativistic and heavy quark
limits. As a result the ratio $f_B/f_D$ is considerably increased
bringing relativistic quark model predictions in agreement with recent
lattice and QCD sum rule calculations. 

 The authors express their gratitude to A. Ali Khan,  
M. M\"uller-Preussker and V.~Savrin  
for support and discussions. Two of us (R.N.F and V.O.G.)
were supported in part by the {\it Deutsche
Forschungsgemeinschaft} under contract Eb 139/2-1, {\it
 Russian Foundation for Fundamental Research} under Grant No.\
00-02-17768 and {\it Russian Ministry of Education} under Grant
No. E00-3.3-45.

\nonumsection{References}


\noindent
\vspace*{-0.6cm}
\begin{thebibliography}{000}
\bibitem{gfm}
V. O. Galkin, A. Yu. Mishurov and R. N. Faustov, {\bibit Sov. J. Nucl. Phys.} 
{\bibbf 53}, 1026 (1991) [{\bibit Yad. Fiz.} {\bibbf 53}, 1676 (1991)].
\bibitem{g}
S. Godfrey, {\bibit Phys. Rev. D} {\bibbf 32}, 189 (1985).
\bibitem{egf}
D. Ebert, V. O. Galkin and R. N. Faustov, {\bibit Phys. Rev. D} 
{\bibbf 57}, 5663 (1998); Erratum  {\bibbf 59}, 019902 (1999).
\bibitem{r}
S. Ryan, {\bibit Nucl. Phys. B (Proc. Suppl.)} {\bibbf 106},
86 (2002).
\bibitem{n}
S. Narison, hep-ph/0202200.
\bibitem{ps}
A. A. Penin and M. Steinhauser, {\bibit Phys. Rev. D} {\bibbf 65},
054006 (2002). 
\bibitem{jl}
M. Jamin and B. O. Lange, {\bibit Phys. Rev. D} {\bibbf 65},
056005 (2002).
\bibitem{pdg}
Particle Data Group, D. E. Groom et al., {\bibit Eur. Phys. J. C}
{\bibbf 15}, 1 (2000).
\bibitem{sr}
S. S\"oldner-Rembold, talk at HEP2001, hep-ex/0109023.
\bibitem{bai}
J. Z. Bai et al. (BES Collaboration) , {\bibit Phys. Lett. B} {\bibbf 429},
188 (1998).
\bibitem{bwd}
C. Bernard et al., {\bibit Phys. Rev. D} {\bibbf 65},
014510 (2002). 
\bibitem{neub1}
M. Neubert, {\bibit Phys. Rev. D} {\bibbf 46}, 1076 (1992).
\end{thebibliography}
\end{document}